\documentclass[a4paper,11pt]{article}
\usepackage{jheppub} 
\title{Differential Representation for Carrollian Correlators}

\abstract{The differential representation of AdS correlators offers a framework to express exchange Witten diagrams as functions of non-local differential operators applied to contact Witten diagrams. In this paper, we develop the differential representation for scalar Carrollian correlators. We first construct this representation using the recently formulated Carrollian limit of AdS Witten diagrams. We then provide an alternate intrinsic analysis that leverages the properties of the Carrollian bulk-to-boundary propagator. Using the differential representation, we also obtain differential Bern-Carrasco-Johansson (BCJ) relations for Carrollian correlators.}

\author[a]{Shankhadeep
Chakrabortty,}
\emailAdd{s.chakrabortty@iitrpr.ac.in}
\affiliation[a]{Department of Physics, Indian Institute of Technology Ropar, Rupnagar, Punjab, India 140001.}
\author[b,c,d]{Subramanya Hegde,}
\emailAdd{subbu@mpp.mpg.de}
\affiliation[b]{Max-Planck-Institut f\"ur Physik, Werner-Heisenberg-Institut, Boltzmannstr. 8, 85748 Garching bei M\"unchen, Germany}
\affiliation[c]{The Institute of Mathematical Sciences, IV Cross Road,CIT Campus, Taramani, Chennai, India 600113.}
\affiliation[d]{Homi Bhabha National Institute, Training School Complex, Anushakti Nagar, Mumbai, India 400085. }

\author [a]{Arpit Maurya}
\emailAdd{arpit.20phz0009@iitrpr.ac.in }
\usepackage[T1]{fontenc}
\usepackage{amsmath,xcolor}
\usepackage{amsmath, amssymb,ascmac,fancybox,mathrsfs,bbold}
\usepackage{color,booktabs,fancyhdr,bm,comment,cancel}
\usepackage{hyperref,appendix}
\usepackage[]{graphicx,graphics}

\makeatletter
\renewcommand{\p@subsection}{}
\makeatother

\makeatletter
\newcommand{\xequal}[2][]{\ext@arrow 0055{\equalfill@}{#1}{#2}}
\def\equalfill@{\arrowfill@\Relbar\Relbar\Relbar}
\makeatother

\renewcommand{\thesection}{\arabic{section}}
\renewcommand{\thesubsection}{\arabic{section}.\arabic{subsection}}

\makeatletter

\@addtoreset{equation}{section} 
\makeatother

\newcommand{\chushi}[1]{ }

\let\calccommentout\iffalse 
\let\calcshow\iftrue 

\newcommand {\mathsym}[1]{{}}
\newcommand {\unicode}[1]{{}}

\allowdisplaybreaks

\begin{document}
\count\footins = 1000 

\maketitle
\newpage
	\section{Introduction}\label{sec-Introduction}
	The AdS (Anti de Sitter)/CFT (Conformal Field Theory) correspondence provides an intriguing connection between theories of quantum gravity in asymptotically AdS spacetimes with field theories in flat space that exhibit conformal invariance \cite{Maldacena:1997re, Witten:1998qj}. This holographic correspondence has,  in the past few decades, provided various insights into gauge and gravitational theories \cite{Ammon:2015wua}. It is therefore of interest to extend this to other spacetimes, in particular to asymptotically flat and de Sitter spacetimes \cite{Barnich:2012aw, Bagchi:2010zz, Strominger:2001pn}. In the context of flat space holography, different approaches have been proposed \cite{Strominger:2013lka,Pasterski:2016qvg,Bagchi:2016bcd,Banerjee:2018gce,Pasterski:2021rjz,Gupta:2021cwo,Raclariu:2021zjz,Pasterski:2021raf,Donnay:2022aba,Donnay:2022wvx,Banerjee:2023bni,Costello:2022jpg,Laddha:2020kvp,Jain:2023fxc,Liu:2024nfc} and 
	the guiding principle has been to develop a formalism that illuminates various features of the infrared triangle between soft theorems, asymptotic symmetries and memory effects \cite{Strominger:2017zoo}. Along this line of research, the reformulation of the flat space S-matrix as boundary correlators in two dimensional (Celestial holography) or three dimensional (Carrollian holography) boundary makes the infrared properties  manifested.
	
	While one can motivate these holographic proposals within the flat space set up, one can also consider the flat space limit of the AdS/CFT correspondence \cite{Susskind:1998vk,Polchinski:1999ry,Giddings:1999qu,Gary:2009ae,Penedones:2010ue} to obtain the S-matrix in flat space.  Recently, this approach has been used to consider particular limits of AdS Witten diagrams that lead to celestial \cite{deGioia:2022fcn} and Carrollian \cite{Bagchi:2023fbj,Bagchi:2023cen,Alday:2024yyj,Bagchi:2024gnn} correlators. In \cite{Bagchi:2023fbj}, by embedding the flat space in the centre of AdS, and by considering a suitable limit one inserts the boundary operators at global AdS time slices $\tau = \pm \frac{\pi}{2}+\frac{u}{L}$, where $L$ is the AdS-radius which is then taken to infinity. This allows one to keep track of the null retarded time $u$ and obtain a Carrollian correlator. Carrollian correlators of different multiplicity and field content have been computed using this approach and they are consistent with the symmetries one expects from a Carrollian field theory. In particular, in \cite{Bagchi:2022emh}, it was argued that scattering amplitudes in modified Mellin basis correspond to Carrollian correlators at the boundary in the {\it delta-function branch}. Computations from the Carrollian limit of AdS Witten diagrams in \cite{Bagchi:2023cen} support this claim. 

 In the context of AdS Witten diagrams, the differential representation arose originally from AdS scattering equations \cite{Eberhardt:2020ewh,Roehrig:2020kck}, and has since then been extensively studied \cite{Armstrong:2020woi,Diwakar:2021juk,Herderschee:2021jbi,Herderschee:2022ntr,Cheung:2022pdk,Armstrong:2022mfr,Li:2022tby,Li:2023azu}. It is a representation of the AdS correlator computed from Witten diagrams in terms of a rational function of differential operators that acts on the contact Witten diagram. Various properties of flat space amplitudes, such as the Bern-Carrasco-Johansson (BCJ) relations \cite{Bern:2008qj}, carry over to AdS correlators as differential equations \cite{Diwakar:2021juk}. Scattering amplitude methods can also be applied then to compute AdS Witten diagrams if one replaces momenta by differential operators and considering all the equations to be acting on the contact Witten diagram \cite{Herderschee:2021jbi}. It is then compelling to consider the Carrollian limit of this representation.

    If we compute Carrollian correlators by using the Carrollian limit of AdS Witten diagrams, we can import various techniques developed for the computation of AdS Witten diagrams. On the other hand, if we compute them as modified Mellin transform of flat space scattering amplitudes one can use various insights from modern scattering amplitude methods. The differential representation of AdS correlators is at the intersection of these two approaches, as it has enabled the connection between amplitude methods and AdS correlators. In this work, we take upon the task of performing the flat space limit of this representation in the context of scalar theories. Remarkably, we will see that the Carrollian limit of the differential representation of AdS correlators tells us that Carrollian correlators are modified Mellin transforms of scattering amplitudes. Our arguments are made using scalar theories, however one can see that they admit a natural generalisation to other theories. We will see that differential representation for Carrollian correlators is extremely simple, as all the non commuting differential operators turn out to be subleading in the large AdS radius expansion, and one is left with commuting differential operators that are translation generators on the boundary.

    The outline of the manuscript is as follows: In section-\ref{sec-Embedding-flat}, we will review the embedding spacetime formalism for AdS/CFT correlators , and the flat space limit of AdS Witten diagrams to Carrollian correlators. In section-\ref{sec-diff-rep-AdS}, we will discuss the differential representation in AdS spacetime. Even though the differential representation arose from AdS ambitwsitor string theory formalism, it was also given a simple explanation in terms of identities satisfied by the AdS bulk-to-boundary propagator. This is the approach we review. To facilitate the Carrollian limit of these identities, we will then write them in global AdS coordinates. This involves a careful coordinate transformation from the embedding space due to additional terms involving the conformal Jacobian factor for the boundary. In section-\ref{diff-rep-Carroll}, we will perform the Carrollian limit of the bulk isometry and boundary conformal generators. In this limit, we will see that the leading order contribution comes from translation generators. This allows us to construct the differential representation for Carrollian correlators. We will use this to illustrate that the Carrollian correlators obtained from the flat space limit of AdS Witten diagrams reproduce modified Mellin transform of the flat space scattering amplitude. We will then perform an intrinsic analysis for the differential representation using the properties of the Carrollian bulk to boundary propagator. In section-\ref{application}, we will show the utility of the differential representation in calculating exchange channel contributions to Carrollian correlators, provided we know the contact diagram contribution. As another application of the differential representation, we will then discuss the color kinematics duality for Carrollian correlators by using differential BCJ relations. In section-\ref{Discussion}, we will summarise our results and outline some future directions.

	\section{Embedding space formalism for AdS correlators and its flat limit}\label{sec-Embedding-flat}
 In this section, we will review the embedding space formalism for AdS correlators \cite{Penedones:2016voo}, and its Carrollian limit that was proposed recently in \cite{Bagchi:2023fbj}. This will help us to understand the AdS differential representation, as well as its Carrollian limit to be discussed in the later sections.
 
 \subsection{Embedding space formalism for AdS correlators}\label{subsec-Embedding}
	In this subsection, we will review the embedding space formalism for AdS space-time as well as for the conformal field theory at the boundary. In particular, we will introduce the isometry generators, bulk-to-boundary and bulk-to-bulk scalar propagators that we need in the later sections. 
 
    The $AdS_{d+1}$ space can be embedded in a $d+2$ dimensional pseudo-Minkowski space coordinatized by $X^0, X^1, X^2, \cdots ,X^{d+1}$ in the following way,
 \begin{eqnarray}
 \label{embedding}
 -{(X^{0})}^2 - {(X^{1})}^2 + \sum_{i=2}^{d+1}{(X^{i})}^2 = - L^2,
 \end{eqnarray}
	where $L$ is the radius of curvature of the $AdS_{d+1}$ spacetime. The metric of the embedding space is described as,
\begin{equation}\label{embedmetric}
	ds^2 = -{(dX^0)}^2  -(dX^{1})^2+ \sum_{i=2}^{d+1}(dX^i)^2 \, .
\end{equation}
The holographic boundary of $AdS_{d+1}$ in embedding formalism is described as the set of outward null rays originated
from the centre of the embedding space $R^{2,d}$. Each point on the null boundary is specified by a $d+2$ dimensional null vector $P^A$ with $A=0,1,\ldots, d+1$, satisfying the following conditions:
\begin{eqnarray}
\label{null}
P^2 = 0, ~~~P^A \sim \lambda P^A,
\end{eqnarray}
where $\lambda > 0$. A set of all such null rays forms a lightcone and the boundary is realized as a specific cross-section $\Sigma$ of the lightcone where each point living on the cross-section corresponds to a null ray. The intrinsic coordinate system of $AdS_{d+1}$ spacetime is directly related to the choice of $\Sigma$. Since later we consider the flat limit of $AdS$ space which is convenient in global $AdS$ coordinate, we implicitly assume that the cross-section is accordingly chosen.  
The advantage of embedding formalism becomes apparent as we identify the group of isometry of Lorentzian $AdS_{d+1}$ is $SO(2,d)$  and the corresponding isometry generators are given as , 
\begin{eqnarray}
\label{isogen}
    D_X^{AB} = X^A \frac{\partial}{\partial X_B} - X^B \frac{\partial}{\partial X_A},
\end{eqnarray}
where $A=0,1,\ldots,d+1$. Using \eqref{isogen}, it is straightforward to construct the Casimir of $AdS$ isometry group acting on a scalar field, 
\begin{eqnarray}
\label{casimir}
    \frac{1}{2}D_{X AB}D_X^{BA}\phi = \Big[ - X^2 {\partial_ X}^2 + X \cdot \partial_X (d+ X \cdot \partial_X)\Big]\phi
\end{eqnarray}
By foliating the embedding space $R^{d,2}$ with $AdS$ surfaces of varying radii $L$, d'Alembert  operator in the embedding space takes the form, 

\begin{equation}
\partial_ X^2 = -\frac{1}{L^{d+1}}\frac{\partial}{\partial L} L^{d+1}\frac{\partial}{\partial L} + \nabla_{AdS_{d+1}}^2
    \label{laplacian}
    \end{equation}
   Moreover, if the embedding equation \eqref{embedding} is taken into account, we get
   \begin{eqnarray}
       \label{relation1}
     X \cdot \partial_X  = L \cdot \partial_L  
   \end{eqnarray}
     By substituting \eqref{laplacian} and \eqref{relation1} back in to the \eqref{casimir}, it is straightforward to show,
   \begin{eqnarray}
      \frac{1}{2}D_{X AB}D_X^{BA}\phi = L^2 \nabla_{AdS}^2 \phi \label{casimir2}
    \end{eqnarray}
    It is important to note that by utilizing the holographic correspondence between the bulk isometry group and the boundary conformal group {$SO(2,d)$}, one can identify the quadratic Casimir in the following way
    \begin{eqnarray}
      \frac{1}{2}D_{X AB}D_X^{BA}\phi = \Delta (\Delta -d) \phi \label{casimir3},
    \end{eqnarray}
    where $\Delta$ is the eigen value of the boundary dilation operator signifying the conformal scaling dimension of the boundary primary field.
    Combining \eqref{casimir2}, \eqref{casimir3} and the Klein-Gordon equation for the massive scalar field in $AdS$ spacetime, $\nabla_{AdS}^2 \phi = m^2 \phi$, leads to the identification,  $m^2 L^2 = \Delta(\Delta - d).$ It is straightforward to show the bulk-to-bulk scalar propagator in Lorentzian $AdS_{d+1}$ 
    obeys the following equation, 
\begin{eqnarray}
    \label{lbulktobulk}
    \Big[\nabla_{AdS_{d+1}}^2 - \frac{\Delta(\Delta - d)}{L^2}\Big] G_{AdS_{d+1}}(X,Y) = i \delta^{(d+1)}(X-Y) 
    \end{eqnarray}
    Equation \eqref{lbulktobulk} can be solved for $G_{AdS_{d+1}}(X,Y)$ by Wick rotating one of the time-like embedding coordinates of the Lorentzian $AdS_{d+1}$  to Euclidean $AdS_{d+1}$ red embedding coordinates describing hyperbolic spacetime, ($H_{d+1}$),
    \begin{eqnarray}
    \label{ebulktobulk}
    \Big[\nabla_{H_{d+1}}^2 - \frac{\Delta(\Delta - d)}{L^2}\Big] G_{H_{d+1}}(X,Y) = -\delta^{(d+1)}(X-Y) 
    \end{eqnarray}
    setting $L=1$, the solution to \eqref{ebulktobulk} yields, 
    \begin{eqnarray} \label{bulkbulkprop}
    G_{H_{d+1}}(X,Y) = \frac{\mathcal{B}_{\Delta}}{\xi^{\Delta} } ~{}_2F_1(\Delta, \Delta - \frac{d}{2} +\frac{1}{2}, 2\Delta - d+1,-\frac{4}{\xi})
    \end{eqnarray}
    where $\xi = {(X-Y)}^2$, $2\Delta = d + \sqrt{d^2 + (2m)^2}$ and 
    \begin{eqnarray}
    \mathcal{B}_{\Delta} = \frac{\Gamma[\Delta]}{2 \pi^{d/2} \Gamma[\Delta-\frac{d}{2} + 1]}
    \end{eqnarray}
 Now, the bulk-to-boundary propagator $E^{H_{d+1}}_{\Delta} (X,P)$ is defined by taking an appropriate limit on $G_{H_{d+1}}(X,Y)$,
 \begin{eqnarray}
 \label{bulkboundprop}
 E^{H_{d+1}}_{\Delta} (X,P) = \lim_{\gamma \to \infty} \gamma^{\Delta}G_{H_{d+1}}(X,Y = \gamma P + \cdots)
 \end{eqnarray}
 Here, $P$ is a null vector satisfying the condition \eqref{null}. Such a limit on  $G_{H_{d+1}}(X,Y)$ and a subsequent Wick rotation together with appropriate $i \epsilon$ prescription yield the desired form of bulk-to-boundary scalar propagator in $AdS_{d+1}$ 
 \begin{eqnarray} \label{bulkbdryprop}
 E_{\Delta} (X,P) = \frac{\mathcal{B}_\Delta}{{(- P\cdot X + i\epsilon) }^\Delta}
 \end{eqnarray}
 Note that once we recover the explicit appearance of $L$, the constant $\mathcal{B}_\Delta$ modifies to 
 \begin{eqnarray}
  \mathcal{B}_{\Delta} = \frac{\Gamma[\Delta]}{ {L^\frac{d-1-2\Delta}{2}} ~2 \pi^{d/2} \Gamma[\Delta-\frac{d}{2} + 1]}
 \end{eqnarray}

  \subsection{Review of flat space limit of AdS correlators}\label{subsec-flat}
In this subsection, we review the flat space limit of AdS bulk-to-boundary correlator by following \cite{Bagchi:2023fbj}. The boundary field theory living on the null boundary of the asymptotically flat spacetime is a co-dimension one theory where the null direction plays a crucial role. 

In order to take a flat space limit of the $AdS$ spacetime, it is convenient to parameterize the embedding spacetime with global $AdS$ coordinates, 
\begin{equation}\label{gAds}
    \begin{split}
        X^0 = -L \dfrac{\cos \tau}{\cos \rho} \, , ~~~ X^{1} =- L \dfrac{\sin \tau}{\cos \rho} \, , ~~~ X^i = L \tan \rho \,  \Omega^i \, , ~~ i= 2, \dots , d+1
    \end{split}
\end{equation}
where $\sum_{i=2}^{d+1}{\Omega^i}^2 =1 $, $\tau\in [-\pi,\pi]$ and $\rho\in [0,\frac{\pi}{2}]$. By substituting \eqref{gAds} in to \eqref{embedmetric}, the 
global $AdS$ metric takes the form,
\begin{equation}\label{eq:adsmetric}
	ds^2 =\frac{L^2}{\cos^2\rho}\left(-d\tau^2+d\rho^2+\sin^2\rho \, d\Omega_{S^{d-1}}^2\right)\, ,
\end{equation}
The boundary of the $AdS$ spacetime is defined as, 
\begin{eqnarray}\label{AdSb}
	P=\lim_{\rho\rightarrow\frac{\pi}{2}}\bigg(\frac{\cos\rho}{L} X \bigg) , ~~~~~ P^2 = 0.
\end{eqnarray}
Combining \eqref{gAds} and \eqref{AdSb}, the boundary points in global $AdS$ parameterization take the following form, 
\begin{equation}\label{AdSbcoord}
    P^0 =- \cos \tau_p \, , ~~~ P^{1} = -\sin \tau_p \, , ~~~ P^i =\Omega^i_p \, ,
\end{equation}
where we use suffix ``$p$'' to represent the boundary value of the global $AdS$ coordinates. 
In order to facilitate the flat limit of $AdS$ spacetime, the re-scaling of global $AdS$ coordinates given by
\begin{equation}\label{coord-rescaling}
\tau=\frac{t}{L},\quad \rho=\frac{r}{L}
\end{equation}
is very effective. Now, $L\to \infty$ limit, a.k.a the flat space limit on the re-scaled coordinates,  yields,
\begin{equation}\label{flatcoord}
	\begin{split}
		X^0 = -L, ~~X^{1}= -t  ,  ~~ X^i = r \Omega^i .
	\end{split}
\end{equation}
By combining \eqref{embedmetric} and \eqref{flatcoord}, the flat metric structure of the $d+1$ dimensional embedded spacetime becomes apparent. 
In order to realize the boundary of the flat geometry as a co-dimension one null boundary $\mathcal{I}^+$, it is customary to express the boundary metric of $AdS$ using the boundary retarded time $u =  L(\tau_p-\pi/2)$ and then take the appropriate flat space limit as $L$ approaches infinity.
\begin{eqnarray}
ds^2_{b} = -\frac{1}{L^2} du^2 + d\Omega_{d-1}^2 \rightarrow - 0. du^2 + d\Omega_{d-1}^2
\end{eqnarray}
Similiar analysis holds for $\mathcal{I}^-$.

One of the significant achievements of embedding formalism of AdS/CFT correspondence is to have a very clear depiction of boundary correlator using the AdS Witten diagram. Likewise, very recently, in Carrollian holography,  such a connection is established between the correlator on the null boundary and flat Witten diagram \cite{Bagchi:2023fbj,Bagchi:2023cen}. Here we review few basic ingredients of such flat Witten diagram, e.g. bulk-to-boundary propagator, vertices, integration measure and internal lines in the context of Carrollian holography. The flat space limit on $AdS$ bulk-to-boundary correlator requires the following, 
\begin{eqnarray}
\label{pdotx}
P \cdot X \approx  u - t +r \Omega \cdot \Omega_p  + \mathcal{O}(\frac{1}{L^2}) 
\end{eqnarray}
Note that the flat limit has to be implemented after taking the dot product of $X^A$ and $P^A$. We can see in \eqref{flatcoord}, that $X^0$ scales as $O(L)$ while from \eqref{AdSbcoord} we can see that $P^0$ scales as $O(L^{-1})$, so that upon taking the dot product they contribute to an $O(1)$ term.

By inserting \eqref{pdotx} in \eqref{bulkbdryprop}, the flat limit of $AdS$ bulk-to-boundary propagator yields, 
 \begin{eqnarray} \label{flatbulkbdryprop}
 K_{\Delta} (x,p) = \frac{\mathcal{B}_\Delta}{{(-u -p\cdot x + i\epsilon) }^\Delta} + \mathcal{O}(\frac{1}{L^2}),
 \end{eqnarray}
where $x= (t, r~\Omega) \in \mathbb{R}^{1,d} $ and $p= (1, ~\Omega_p) \in \mathbb{R}^{1,d}$.
In deriving \eqref{flatbulkbdryprop}, it is assumed that $ \tau_p  = \frac{u}{L} +\pi/2$ and the bulk-to-boundary propagator corresponds to the outgoing mode. Similarly for $\tau_p = \frac{u}{L} - \pi/2$, the bulk-to-boundary propagator for the outgoing mode becomes, 
\begin{eqnarray} \label{flatbulkbdrypropout}
 K_{\Delta} (x,p) = \frac{\mathcal{B}_\Delta}{{(u + p\cdot x + i\epsilon) }^\Delta}  + \mathcal{O}(\frac{1}{L^2})
 \end{eqnarray}
Another important element to introduce the flat Witten diagram is the interaction vertex. Such a flat vertex operator can be obtained by taking a flat limit on the interaction vertex in $AdS$ spacetime for non-derivative couplings.
\begin{equation}
\label{vertex}
	i ~\lambda \int \sqrt{-{g_{AdS}}_{d+1}}~ d\tau ~d\rho ~d\Omega_{d-1} = i ~\lambda \int \sqrt{-{g_{R^{1,d}}}}~ dt ~dr ~d\Omega_{d-1} + \mathcal{O}(\frac{1}{L^2}) .
\end{equation}
Finally, the internal line of flat Witten diagram requires the knowledge of flat version of bulk-to-bulk propagator. It turns out that a consistent flat limit on \eqref{lbulktobulk} results into the corresponding time orderd Green's function in flat spacetime.
\begin{eqnarray}
    \label{flatlbulktobulk}
    \Big[\nabla_{\mathbb{R}^{1,d}}^2 - m^2 \Big] G_{\mathbb{R}^{1,d}}(x,y) = i \delta_{\mathbb{R}^{1,d}}(x-y) ,
    \end{eqnarray}
    where $G_{\mathbb{R}^{1,d}}(x,y)$ is the Feynman propagator, and we have taken $\Delta \sim O(L)$ and defined $\frac{\Delta}{L}=m$.
 Having introduced all the elements of flat Witten diagrams in Carrollian holography, in the next section we review the differential representation in AdS. 
  \section{Differential Representation in
 AdS holography}
 \label{sec-diff-rep-AdS}
 
 In this section, we will start with a discussion on differential representation for AdS correlators. We will review the construction of the differential representation by using some identities satisfied by the bulk-to-boundary propagator. We will rewrite these identities in global AdS coordinates to ease the presentation of their Carrollian limit needed for the later sections.

\subsection{Differential representation in embedding spacetime}\label{subsec-diff-rep-Embedding}
\begin{figure}[h]
\centering
\includegraphics[width=0.32\linewidth]{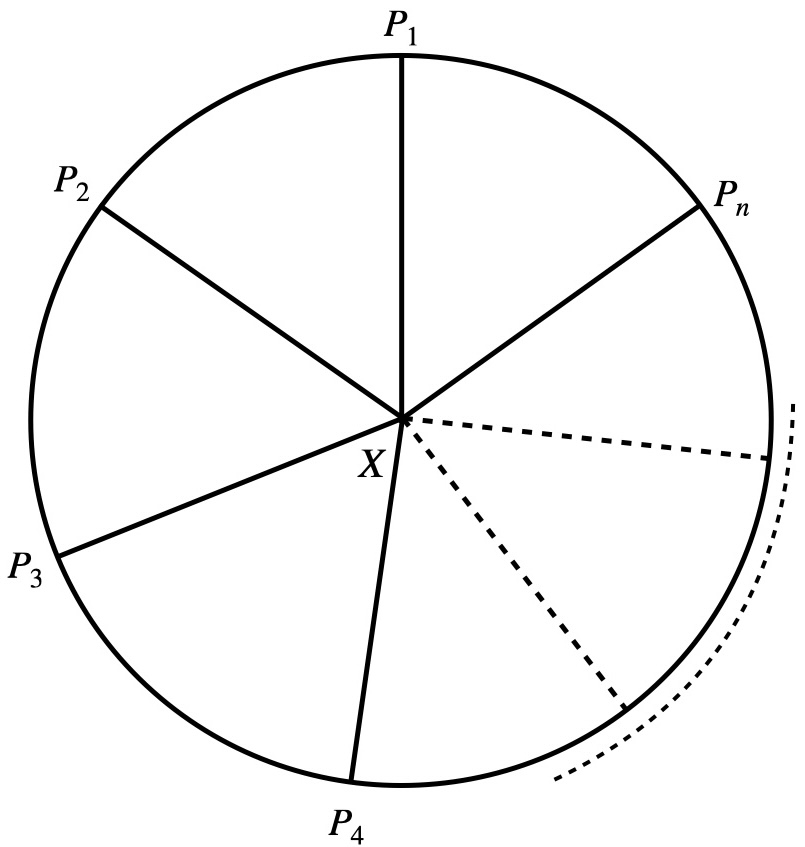}
\caption{n-point contact Witten diagram}
\label{fig:1}
\end{figure}

 Here, we review the differential representation for AdS correlators in embedding formalism \cite{Eberhardt:2020ewh}. In general, the differential representation of a n-point correlator takes the following form
\begin{equation}
    \mathcal{A}_n=\hat{\mathcal{A}}_n C_n
\end{equation}
where $\hat{\mathcal{A}}_n$ is the differential operator and $C_n$ represents the n-point contact Witten diagram in Fig.(\ref{fig:1}), mathematically expressed as
\begin{eqnarray}
    C_n=\int_{AdS} dX \prod_i^n E(P_i,X)
\end{eqnarray}
where $E(P_i,X)$ is the AdS bulk-to-boundary propagator connecting the bulk point $X$ to the boundary point $P_i$ in embedding coordinates, as given in \eqref{bulkbdryprop}.
The scalar differential operator $\hat{\mathcal{A}}_n$ can solely be  written in terms of conformal generators given by 
 \begin{equation} \label{confbdryprop}
         D_{P}^{AB}= P^A \frac{\partial}{\partial P_B}-P^B \frac{\partial}{\partial P_A}.
         \end{equation}
A crucial ingredient in deriving the differential representation is an identity for the bulk-to-boundary propagator that reads,
\begin{eqnarray}\label{singleadsidentity}
	\bigg(D_X^{AB}+D_P^{AB}\bigg) E_{\Delta}(P,X)=0.
\end{eqnarray}
One can derive this identity from the transformation of the bulk-to-boundary propagator upon the simultaneous action of a bulk isometry transformation and a corresponding boundary conformal transformation. Under a bulk isometry transformation, $X^{\prime AB}= \left(e^{\frac{1}{2}\lambda_{CD}D_X^{CD}}\right)X^{AB}$, and the corresponding boundary conformal transformation, $P^{\prime AB}=\left(e^{\frac{1}{2}\lambda_{CD}D_P^{CD}}\right)P^{AB}$, the bulk-to-boundary propagator transforms as,
\begin{equation}\label{trans-bulk-to-boundary}
E_\Delta(P^\prime, X^\prime)=J^{-\Delta} E_\Delta(P, X),
\end{equation}
where $J$ is the Jacobian factor that arises due to the conformal transformation on the boundary. In embedding coordinates, the action of the $SO(d,2)$ conformal symmetry is linearised and the Jacobian is the determinant of an $SO(d,2)$ matrix, and hence $J=1$. By considering an infinitesimal transformation, we obtain the identity \eqref{singleadsidentity}. It is also important to note that while the Jacobian above is one, the scaling dimension of the bulk-to-boundary propagator is manifested in embedding coordinates using the projective scaling on the embedding light cone, $E_\Delta(\lambda P,X)=\lambda^{-\Delta}E_\Delta(P, X)$. When we go to a different coordinate system such as the global AdS coordinates, we choose a specific section of the projective light cone, and the scaling dimension will be encoded via a non-trivial Jacobian. We will return to this in the next section.

\begin{figure}[h]
\centering
\includegraphics[width=0.32\linewidth]{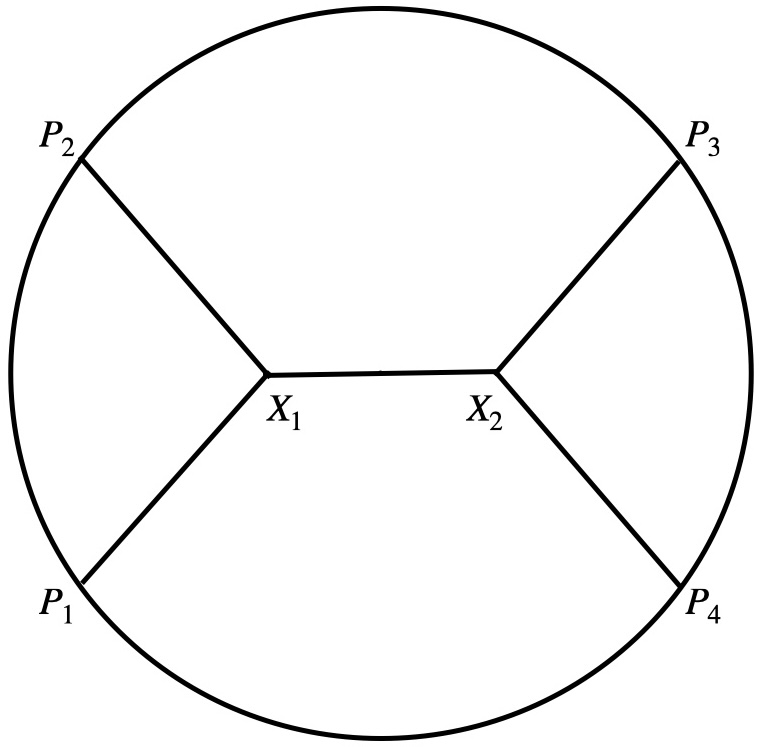}
\caption{s-channel Witten diagram}
\label{fig:2}
\end{figure}

When we have a product of multiple bulk-to-boundary propagators supported on the same bulk point, the identity above trivially generalises to,
\begin{eqnarray} \label{adsidentity}
    \bigg(D_X^{AB}+\sum_i D_{P_i}^{AB}\bigg) \prod_i E(P_i,X)=0
\end{eqnarray}
One can easily verify this identity using \eqref{isogen}, \eqref{bulkbdryprop} and \eqref{confbdryprop}. We will illustrate the discussion above by using s-channel Witten diagram for scalar fields as an example. An s-channel Witten diagram Fig.\ref{fig:2} is defined as
\begin{eqnarray} \label{schannel}
    \mathcal{A}_{s}=\int_{AdS} dX_1dX_2 E_{\Delta_3}(P_3,X_1)E_{\Delta_4}(P_4,X_1)G_{\Delta}(X_1,X_2)E_{\Delta_1}(P_1,X_2)E_{\Delta_2}(P_2,X_2)
\end{eqnarray}

\begin{figure}[h]
\centering
\includegraphics[width=0.32\linewidth]{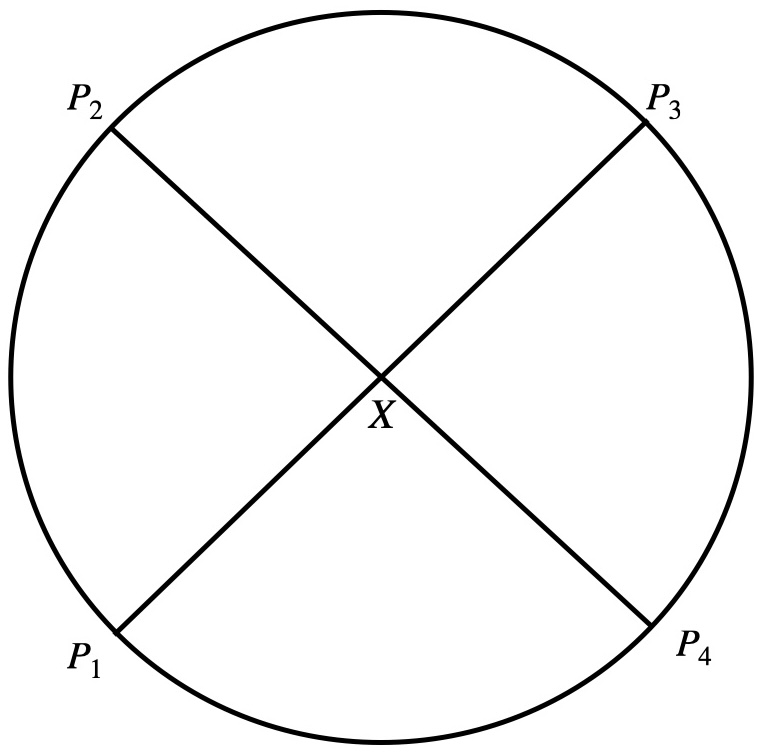}
\caption{ 4-point contact Witten diagram}
\label{fig:3}
\end{figure}

We can now replace the Green's function $G_{\Delta}(X_1,X_2)$ by inverse of free field operator acting of delta function given by
\begin{equation}
    G_{\Delta}(X_1,X_2)= - \frac{1}{\nabla_{X_1}^2-\Delta(\Delta-d)} \delta(X_1-X_2)
\end{equation}
where the AdS Laplacian $\nabla_{X_1}^2$ can be written in terms of AdS isometry generators using \eqref{casimir2}. Now, \eqref{schannel} can be rewritten as 
\begin{equation} \label{schannel1}
    \mathcal{A}_{s}=\int_{AdS} dX_1dX_2 E_{\Delta_3}(P_3,X_2)E_{\Delta_4}(P_4,X_2) \frac{(-\delta(X_1-X_2))}{\nabla_{X_1}^2-\Delta(\Delta-d)}E_{\Delta_1}(P_1,X_1)E_{\Delta_2}(P_2,X_1)
\end{equation}
Using the identity \eqref{adsidentity}, we can replace the $D_{X_1}^2$ with scalar products of the conformal boundary generators given by,
\begin{eqnarray}
    D_{12}^2=(D_{P_1,AB}+D_{P_2,AB})(D_{P_1}^{AB}+D_{P_2}^{AB})
\end{eqnarray}
Then, \eqref{schannel1} simplifies as,
\begin{eqnarray}
    \mathcal{A}_{s}= \frac{1}{D_{12}^2-\Delta(\Delta-d)}\int_{AdS} dX \  \prod_{i=1}^{4}E_{\Delta_i}(P_i,X)= \frac{1}{D_{12}^2-\Delta(\Delta-d)} C_4
\end{eqnarray}
where $C_4$ is 4-point contact Witten diagram Fig.\ref{fig:3}.

\subsection{Differential representation in global AdS coordinates}\label{subsec-diff-rep-globalAdS}
In the previous section, using the differential representation, we discussed how a tree level boundary correlator is obtained from a contact Witten diagram upon the action of non-local operators. These non-local operators are written in terms of conformal generators of the  boundary spacetime. In this section, we will choose $d=3$ and we will review the bulk isometry generators and the conformal boundary generators  of $\mathrm{AdS}_4$ in global coordinates. Setting up the differential representation in global AdS coordinates is an important step for the later sections where we take the Carrollian limit of AdS correlators. Even though we are only doing a coordinate transformation of the identity for the bulk-to-boundary propagator given in \eqref{adsidentity}, we need to take special care in going from the boundary coordinates in embedding space formalism to the boundary coordinates that we obtain from the boundary limit of global AdS coordinates, as the boundary of AdS is encoded in the embedding light cone via a specific cross-section.

	Here, we shall explicitly reproduce all the isometry generators of AdS$_4$ in the global coordinates. In embedding coordinate, the bulk isometry generators are given by \eqref{isogen}. Using the transformation rule \eqref{gAds}, we can write the the isomery generators of AdS$_4$ in global coordinates. From \eqref{eq:adsmetric}, we can straight forwardly see that $\tau$-translation and rotations along $S^2$ are isometries of AdS. In fact, they correspond to the generators,
	\begin{align}\label{isometry-global-AdS-1}
	D_X^{01}&=-\frac{\partial}{\partial \tau},\nonumber\\
	D_X^{23}&= \frac{\partial}{\partial \phi},\nonumber\\
	D_X^{24}&=-\cos\phi \frac{\partial}{\partial \theta }+\cot\theta \sin\phi \frac{\partial}{\partial\phi},\nonumber\\
	D_X^{34}&=-\sin\phi \frac{\partial}{\partial \theta }-\cot\theta  \cos\phi  \frac{\partial }{\partial \phi }.
	\end{align}
	The rest of the isometry generators turn out to be,
	\begin{align}\label{isometry-global-AdS-2}
	D_X^{02}&=  \sin \theta \cos \phi \left( \sin \rho \sin \tau	\frac{\partial }{\partial \tau}
	-  \cos \rho \cos \tau \frac{\partial }{\partial \rho}\right)-\cos\theta \cos\phi \cos \tau\csc\rho \frac{\partial }{\partial \theta }\nonumber\\
	&+\sin\phi \cos \tau\csc\theta \csc \rho\frac{\partial }{\partial \phi },\nonumber\\
	D_X^{03}&=\sin \theta \sin \phi \left( \sin \rho \sin \tau\frac{\partial }{\partial \tau}- \cos \rho \cos \tau\frac{\partial }{\partial \rho}\right)-\cos \theta \sin \phi \cos \tau\csc \rho \frac{\partial }{\partial \theta }\nonumber\\
	&-\cos \phi \cos \tau\csc\theta \csc \rho\frac{\partial }{\partial \phi } \nonumber\\
	D_X^{04}&= \cos \theta  \left(\sin \rho \sin \tau \frac{\partial }{\partial \tau}- \cos \rho\cos \tau \frac{\partial}{\partial \rho}\right)+\sin \theta  \cos \tau\csc\rho\frac{\partial }{\partial \theta },\nonumber\\
	D_X^{12}&=- \sin\theta \cos\phi \left(\sin \rho \cos\tau\frac{\partial }{\partial \tau}+\cos \rho \sin \tau\frac{\partial }{\partial \rho}\right)-\cos\theta \cos\phi \sin \tau\csc \rho\frac{\partial }{\partial \theta }\nonumber \\ 
	&+\sin \phi  \sin \tau\csc\theta \csc \rho\frac{\partial }{\partial \phi },\nonumber\\
	D_X^{13}&=- \sin\theta \sin \phi \left( \sin \rho \cos \tau\frac{\partial}{\partial \tau}+ \cos \rho \sin \tau\frac{\partial }{\partial r}\right)-\cos\theta \sin\phi   \sin \tau\csc \rho\frac{\partial }{\partial \theta } \nonumber\\
	&-\cos\phi \sin \tau \csc\theta \csc \rho\frac{\partial }{\partial \phi },\nonumber\\
	D_X^{14}&=- \cos \theta \left( \sin \rho \cos \tau\frac{\partial }{\partial \tau} + \cos \rho \sin \tau \frac{\partial }{\partial \rho}\right)+\sin\theta \sin \tau\csc \rho\frac{\partial }{\partial \theta }
	\end{align}
	While in terms of $\tau$ and $\rho$ coordinates, all the generators above seem to be the same order in $L$, in terms of the rescaled coordinates $t=L\tau, r=L\rho$, some isometry generators dominate over others. This will be made explicit in the next section when we write the flat limit of these generators.

    Similarly, the conformal generators on the boundary given in embedding coordinates in \eqref{confbdryprop} can be transformed to global AdS coordinates using \eqref{AdSbcoord}. Clearly, time translation and rotations along $S^2$ remain symmetries on the boundary. The corresponding generators are given as,
   	\begin{align}\label{AdS-boundary-gen-1}
   	D_P^{01}&=-\frac{\partial}{\partial \tau_p},\nonumber\\
   	D_P^{23}&= \frac{\partial}{\partial \phi_p},\nonumber\\
   	D_P^{24}&=-\cos\phi_p \frac{\partial}{\partial \theta_p}+\cot\theta_p \sin\phi_p \frac{\partial}{\partial\phi_p},\nonumber\\
   	D_P^{34}&=-\sin\phi \frac{\partial}{\partial \theta_p }-\cot\theta_p  \cos\phi_p  \frac{\partial }{\partial \phi_p }.
   \end{align} 
   The boundary of global AdS$_4$ is $\mathbb{R} \times S^2$. The generators above encode the isometries of $\mathbb{R}$ and $S^2$ separately. The rest of the conformal generators are given by,    
   \begin{align}\label{AdS-boundary-gen-2}
   	D_P^{02}&=\sin \theta_p \sin \tau_p \cos \phi_p \frac{\partial }{\partial \tau_p}-\cos \theta_p  \cos \tau_p \cos\phi_p \frac{\partial }{\partial \theta_p}+\cos \tau_p \sin \phi_p \csc \theta_p\frac{\partial }{\partial \phi_p},\nonumber\\
   	D_P^{03}&=\sin \theta_p \sin \tau_p \sin \phi_p \frac{\partial }{\partial \tau_p}-\cos \theta_p \cos \tau_p \sin \phi_p \frac{\partial }{\partial \theta_p}-\cos \tau_p \cos \phi_p \csc \theta_p\frac{\partial }{\partial \phi_p},\nonumber\\
   	D_P^{04}&=\cos \theta_p \sin \tau_p \frac{\partial }{\partial \tau_p}+\sin \theta_p \cos \tau_p \frac{\partial }{\partial \theta_p},\nonumber\\
   	D_P^{12}&=-\sin \theta_p \cos \tau_p \cos \phi_p \frac{\partial }{\partial \tau_p}-\cos \theta_p \sin \tau_p \cos\phi_p \frac{\partial }{\partial \theta_p}+\sin \tau_p \sin \phi_p \csc \theta_p\frac{\partial }{\partial \phi_p},\nonumber\\
   	D_P^{13}&=-\sin \theta_p \cos \tau_p \sin \phi_p \frac{\partial }{\partial \tau_p}-\cos \theta_p \sin \tau_p \sin \phi_p \frac{\partial }{\partial \theta_p}-\sin \tau_p \cos \phi_p \csc \theta_p\frac{\partial }{\partial \phi_p}\nonumber\\
   	D_P^{14}&=-\cos \theta_p \cos \tau_p \frac{\partial }{\partial \tau_p}+\sin \theta_p \sin \tau_p \frac{\partial }{\partial \theta_p}.
   \end{align}
	So far, all we have done is a coordinate transformation that needed a simple application of chain rule of differentiation on the generators. While this is straightforward, a technical subtelty we would like to highlight in this section is that since the boundary coordinates above represent a particular cross-section of the projective light cone in the embedding space, we need to be careful in converting the identity in \eqref{adsidentity} in global AdS coordinates. If we write the action of the transformed $D_X^{AB}, D_P^{AB}$ operators on the transformed bulk-to-boundary propagators to transform the identity, we miss the contribution of the non-trivial Jacobian factor that we briefly mentioned in the previous section. To recover this factor, note that when we act the transformed differential generators on the bulk-to-boundary propagator in global AdS coordinates, the answer is proportional to the bulk-to-boundary propagator. The proportionality factor is the Jacobian. In any case, for the generators that correspond to time translation and rotations on $S^2$, the Jacobian $J=1$ since they are isometries of both the bulk and the boundary. Therefore, 
	\begin{align}\label{noscale-identity-AdS}
		\left( D_X^{AB}+\sum_iD_{P_i}^{AB} \right)\prod_i E_{\Delta_i}(X,P_i)&=0, \, \{AB\}=01, 23, 24, 34.
	\end{align}
	For other generators, we get,
	\begin{align}\label{scale-identity-AdS}
		\left( D_X^{AB}+\sum_iD_{P_i}^{AB}+\sum_i \Delta_i\xi^{AB}(P_i) \right)\prod_i E_{\Delta_i}(X,P_i)&=0, \, \{AB\}\neq 01, 23, 24, 34,
	\end{align}
	where $J\equiv 1+\frac{1}{2}\lambda_{AB}\xi^{AB}$ defines the factor $\xi^{AB}$. We can use the conformal Killing equations to determine this factor. We will write a given conformal generator $D_P^{AB}$ as,
	\begin{equation}
	D_{P}^{AB}\equiv\xi^{ABa}(P)\frac{\partial}{\partial x_p^a},
	\end{equation}
	where $x_p^a=\tau_p,\theta_p,\phi_p$ are the coordinates on the boundary. Then the conformal Killing equation reads,
	\begin{equation}
	\mathcal{L}_\xi g_{ab}(P)=2g_{ab}(P)\xi^{AB}(P),
	\end{equation}
	where the normalisation on the RHS manifests that the metric has scaling dimension $2$. Therefore, we get,
	\begin{equation}
	\xi^{AB}(P)=\frac{1}{3}\nabla_a\xi^{ABa}(P).
	\end{equation}
	In Table-\ref{table:scale-factors}, we have given the scale factors for various generators explicitly.
	\begin{table}
	\small
	\centering	
	    \begin{tabular}{|c|c|
		}
		\hline
		$AB$ & $\xi^{AB}(P)$\\
		\hline
		$02$ & $\cos \tau_{p} \sin \theta_{p}  \cos \phi_{p}$\\
		\hline
		$03$ & $ \cos \tau_{p} \sin \theta_{p}  \sin \phi_{p}$\\
		\hline
		$04$ & $\cos \tau_{p} \cos \theta_{p} $\\
		\hline
		$12$ & $ \sin \tau_{p} \sin \theta_{p} \cos \phi_{p}$\\
		\hline
		$13$ & $\sin \tau_{p}  \sin \theta_{p} \sin \phi_{p}$\\
		\hline
		$14$ & $ \sin \tau_{p} \cos \theta_{p}$\\
		\hline
	\end{tabular}
	\caption{Scale factors for boundary conformal transformations.}
	\label{table:scale-factors}
	\end{table}
	In the next section, we will perform the flat space limit of the identities \eqref{noscale-identity-AdS} and \eqref{scale-identity-AdS} using the Carrollian limit proposed in \cite{Bagchi:2023fbj} to obtain the differential representation for the Carrollian correlators.
 
 \section{Differential Representation in
 Carrollian holography}\label{diff-rep-Carroll}
 In this section, we will construct the differential representation for Carrollian correlators. We will first do this by taking the flat space limit of the AdS differential representation. We will then show an independent and simple derivation based on the definition of Carrollian conformal primary wave function.
\subsection{Carrollian limit of the conformal generators}
We will now consider the flat space limit of the differential isometry generators of AdS and the conformal generators on its boundary. To consider the flat space limit of the isometry generators in \eqref{isometry-global-AdS-1} and \eqref{isometry-global-AdS-2}, we will first rescale the coordinates according to \eqref{coord-rescaling}, and then take the $L \rightarrow \infty$ limit. In this limit, some generators scale as $L$ whereas other generators scale as $O(1)$. 

The generators that scale as $L$ are,
\begin{align}
	D_x^{01} &=-L \frac{\partial}{\partial t}, \nonumber\\
	D_x^{02} & = 
-L \bigg( \sin\theta \cos\phi  \frac{\partial }{\partial r}
+\frac{1}{r} \cos\theta \cos\phi \frac{\partial}{\partial\theta}- \frac{1}{r} \sin\phi\csc\theta\frac{\partial}{\partial\phi} \bigg)=-L\frac{\partial}{\partial x},\nonumber\\
	D_x^{03} &= -L \bigg( \sin\theta \sin\phi  \frac{\partial }{\partial r}
+\frac{1}{r} \cos\theta \sin\phi \frac{\partial}{\partial\theta}+ \frac{1}{r} \cos\phi\csc\theta\frac{\partial}{\partial\phi} \bigg)=-L\frac{\partial}{\partial y},\nonumber\\
	D_x^{04}&= -L \bigg( \cos\theta  \frac{\partial }{\partial r}
-\frac{1}{r} \sin\theta \frac{\partial}{\partial\theta} \bigg)=-L\frac{\partial}{\partial z},
 \end{align}
 where in the last equality on each line we have gone from spherical polar coordinates to Cartesian coordinates in the spatial directions. We see above that the generators that are leading order in $L$ are the translation generators in Minkowski space-time. We would also expect rotations and boosts. The rotation generators are,
 \begin{align}
 	D_x^{23}&= \frac{\partial}{\partial \phi},\nonumber\\
 	D_x^{24}&=-\cos\phi \frac{\partial}{\partial \theta }+\cot\theta \sin\phi \frac{\partial}{\partial\phi},\nonumber\\
 	D_x^{34}&=-\sin\phi \frac{\partial}{\partial \theta }-\cot\theta  \cos\phi  \frac{\partial }{\partial \phi },
 \end{align}
 and the boost generators are,
 \begin{align}
  D_x^{12}&= -t \sin{\theta} \cos\phi 
 \frac{\partial}{\partial r}-\frac{t}{r} \cos\theta \cos\phi \frac{\partial}{\partial\theta}-r \sin\theta \cos\phi \frac{\partial}{\partial t}+\frac{t}{r}\sin\phi\csc\theta\frac{\partial}{\partial \phi}\\
	D_x^{13}&= -t \sin{\theta} \sin\phi 
 \frac{\partial}{\partial r}-\frac{t}{r} \cos\theta \sin\phi \frac{\partial}{\partial\theta}-r \sin\theta \sin\phi \frac{\partial}{\partial t}-\frac{t}{r}\cos\phi\csc\theta\frac{\partial}{\partial \phi}\\
	D_x^{14}&= -t \cos{\theta}
 \frac{\partial}{\partial r}+\frac{t}{r} \sin\theta \frac{\partial}{\partial\theta}-r \cos\theta  \frac{\partial}{\partial t}
 \end{align}
 What we have done above is essentially an In\"on\"u-Wigner contraction from AdS to Minkowski isometry generators \cite{Giddings:1999jq,Goncalves:2014rfa}. 
 
 More non-trivially, we can now perform the same on the conformal boundary generators.  To consider the Carrollian limit of \eqref{AdS-boundary-gen-1} and \eqref{AdS-boundary-gen-2}, we will first redefine the boundary coordinate $\tau_p$ according to,
 \begin{equation}
 \tau_p=\frac{\pi}{2}+\frac{u}{L},
 \end{equation}
 and then take the $L\rightarrow \infty$ limit. Leading generators are again,
 \begin{align}\label{flat-boundary-translation}
 D_p^{01}&=-L \frac{\partial}{\partial u},\nonumber\\
 D_p^{02}&= L \sin\theta_p \cos\phi_p \frac{\partial}{\partial u},\nonumber\\
 D_p^{03}&= L \sin\theta_p \sin\phi_p \frac{\partial}{\partial u},\nonumber\\
 D_p^{04}&= L \cos\theta_p \frac{\partial}{\partial u},
 \end{align}
 these are nothing but the translation generators on the boundary. The rotation generators are,
 \begin{align}
 	D_p^{23}&= \frac{\partial}{\partial \phi_p},\nonumber\\
 	D_p^{24}&=-\cos\phi_p \frac{\partial}{\partial \theta_p }+\cot\theta_p \sin\phi_p \frac{\partial}{\partial\phi_p},\nonumber\\
 	D_p^{34}&=-\sin\phi_p \frac{\partial}{\partial \theta_p }-\cot\theta_p  \cos\phi_p  \frac{\partial }{\partial \phi_p },
 \end{align}
and the boost generators are, 
\begin{align}
D_p^{12}&= - \cos\theta_p \cos\phi_p \frac{\partial}{\partial\theta_p}+u \sin\theta_p \cos\phi_p \frac{\partial}{\partial u}+\sin\phi_p\csc\theta_p\frac{\partial}{\partial \phi_p},\nonumber\\
D_p^{13}&= - \cos\theta_p \sin\phi_p \frac{\partial}{\partial\theta_p}+u \sin\theta_p \sin\phi_p \frac{\partial}{\partial u}-\cos\phi_p\csc\theta_p\frac{\partial}{\partial \phi_p},\nonumber\\
D_p^{14}&= \sin\theta_p \frac{\partial}{\partial\theta_p}+u \cos\theta_p  \frac{\partial}{\partial u}.
\end{align} 
Note that the flat limit of the boundary conformal generators is not trivial. For instance, if we were to do it for the celestial case following the prescription of \cite{deGioia:2022fcn}, then we will have fixed $\tau_p=\frac{\pi}{2}$, and we would have missed the boundary translation generators in \eqref{flat-boundary-translation}. While these are not expected to be the symmetries of the celestial sphere, these will play an important role in the differential representation, as translation generators are tied to the flat space Laplacian operator. We will provide a prescription for the celestial case in the coming sections, by using an intrinsic analysis.

\subsection{Differential Representation using conformal Carrollian generators}
We will now consider the flat space version of \eqref{adsidentity} to construct the differential representation for Carrollian correlators. The Carrollian bulk-to-boundary propagator from \eqref{flatbulkbdryprop} satisfies the following relations with respect to the flat space isometry generators and the boundary Carrollian conformal generators discussed in the previous subsection.
\begin{equation}\label{Carrolian-conformal-identity}
	\left( D_x^{AB}+\sum_i D_{p_i}^{AB} \right) \prod _i K_{\Delta_i}(x,p_i)=0, \, \{AB\}=01, 02, 03, 04, 23, 24, 34.
\end{equation}
The scaling factors for $\{AB\}=02,03,04$ from Table-\ref{table:scale-factors} are subleading in the flat space limit as $L \rightarrow \infty$, resulting in the relation above for the leading order Carrollian bulk-to-boundary propagator given in \eqref{flatbulkbdryprop}. For the rest of the generators, one has scaling dimensions appearing on the R.H.S. along with a scale factor. This is expected on general symmetry grounds as the generators below correspond to Lorentz boosts; and the Carrollian bulk-to-boundary propagator, which is nothing but the Carrollian conformal primary wave function, is a boost eigenfunction. The relations are given as,
\begin{equation}
\left( D_x^{AB}+\sum_i D_{p_i}^{AB} \right)\prod_i K_{\Delta_i}(x,p_i)=-\left(\sum_i\Delta_i \xi^{AB}(p_i)\right)\prod_i K_{\Delta_i}(x,p_i),\, \{AB\}=12,13,14,
\end{equation}
where the scale factors are as given in Table-\ref{table:Carrollian-scale-factors} .
	\begin{table}
	\small
	\centering	
	\begin{tabular}{|c|c|
		}
		\hline
		$AB$ & $\xi^{AB}(p)$\\
		\hline
		$12$ & $\sin \theta_{p} \cos \phi_{p}$\\
		\hline
		$13$ & $  \sin \theta_{p} \sin \phi_{p}$\\
		\hline
		$14$ & $ \cos \theta_{p}$\\
		\hline
	\end{tabular}
	\caption{Scale factors for boundary Carrollian conformal transformations.}
	\label{table:Carrollian-scale-factors}
\end{table}

For studying the flat space limit of the differential representation, it is important to keep track of the order of scaling of the differential generators as $L \rightarrow \infty$. In particular, the AdS Laplacian goes to the flat space Laplacian as,
\begin{equation}
\nabla_{AdS}^2\rightarrow \nabla_{\mathbb{R}^{1,3}}^2+O\left(\frac{1}{L^2}\right) ~~,
\end{equation}
and in terms of the differential generators this reads,
\begin{align}
\sum_{A < B; \,A, B=0}^4 D_X^{AB}D_{X,AB}\rightarrow &\left(D_x^{01}D_x^{01}-D_x^{02}D_x^{02}-D_x^{03}D_x^{03}-D_x^{04}D_x^{04}\right)+O(1)\nonumber\\
=&-L^2(\partial_\mu\partial^\mu)+O(1).
 \end{align}
 i.e., the translation generators dominate as $L\rightarrow \infty$ giving rise to the flat space Laplacian as expected. When the Laplacian acts on a product of bulk-to-boundary propagators, we can replace the bulk differential operators with boundary conformal generators using \eqref{Carrolian-conformal-identity}. Namely, we get from the cases $\{AB\}=01,02,03,04$, 
 \begin{equation}
 \left(\frac{\partial}{\partial x^\mu}-\sum_i\eta_{\mu\nu}\tilde{q}_i^\nu\frac{\partial}{\partial u_i}\right)\prod_i K_{\Delta_i}(x,p_i)=0,
 \end{equation}
 where,
 \begin{equation} \label{tilde-q}
 \tilde{q}_i^\mu=(1,\sin\theta_{pi}\cos\phi_{pi},\sin\theta_{pi}\sin\phi_{pi},\cos\theta_{pi}).
 \end{equation}
 Therefore,
 \begin{equation}
 \left[\nabla_{\mathbb{R}^{1,3}}^2-\bigg(\sum_i\tilde{q}_{i}\frac{\partial}{\partial u_i}\bigg)^2\right]\prod_iK_{\Delta_i}(x,p_i)=0.
 \end{equation}
 The above equation holds intrinsically in Carrollian holography, which we will discuss in the next section. When we consider the flat space limit of the AdS Witten diagram formalism, then the above is the leading term in the $L\rightarrow\infty$ limit.
 
 Following the above analysis, we consider the similar technique as presented previously in the section \ref{subsec-diff-rep-Embedding} and re-express a Carrollian conformal correlator, with exchange channel contributions, in terms of the contact diagram by using,
 \begin{equation}\label{flat-diff-rep}
 	\mathcal{A}_n(\{p_i\})=\hat{\mathcal{A}}_n(\{D_{p_i\mu}\}) C_n(\{p_i\}),
 \end{equation}
 where,
 \begin{equation}
 D_{p_i\mu}\equiv\eta_{\mu\nu}\tilde{q}_i^\nu\frac{\partial}{\partial u_i},
 \end{equation}
 are the differential operators in terms of which $\hat{\mathcal{A}}_n$ resembles a flat space scattering amplitude with,
 \begin{equation}
 p_{i\mu}\rightarrow D_{p_i\mu}.
 \end{equation}
 Note that, since $\tilde{q}_i^2=0$,
 \begin{equation}
 D_{p_i}^2=0.
 \end{equation}
 Also, the operators $D_{p_i}^\mu$ all commute, since they are all simple partial derivatives and $\tilde{q}_{i\mu}$ has no dependence on $u_i$.
 \begin{equation}
 [D_{p_i\mu},D_{p_j\nu}]=0, \,\forall i,j,\mu,\nu.
 \end{equation}
 This has to be contrasted with the AdS/CFT case, where the differential generators don't commute. In our case, the commutation is simply the reflection of the fact that the translation group is abelian and the rest of the Carrollian conformal generators are subleading in the $L\rightarrow \infty$ limit, and therefore do not contribute to the Laplacian. 

  In the next section, we will see that Carrollian correlators computed using the flat space limit of AdS Witten diagrams match modified Mellin transforms of scattering amplitudes. While this can be seen using several approaches \cite{Bagchi:2023cen, Alday:2024yyj}, we use the differential representation approach. We will further see that the boundary differential operators $D_{p_i\mu}$ defined above become momenta when they act on the kernel of the modified Mellin transform. Thus, the modified Mellin basis will be shown to simultaneously diagnonalise these operators.

 \subsection{Carrollian amplitude and Witten diagrams}
 In recent literature, there are two methods that one could follow to construct Carrollian correlators in flat space holography:
 \begin{itemize}
 	\item Carrollian correlators are obtained by flat space limit of AdS Witten diagrams \cite{Bagchi:2023fbj}. As we reviewed in section-\ref{subsec-flat}, in this point of view, one takes the Carrollian limit of the building blocks of Witten diagrams.
 	\item Carrollian correlators are obtained by taking the modified Mellin transform of the flat space scattering amplitude \cite{Bagchi:2022emh}. In this framework, one transforms the position space scattering amplitude in flat spacetime to a boost eigenbasis by using Carrollian conformal primary wave functions \cite{Banerjee:2018gce}.
 \end{itemize}    
 The connection between these two approaches was made in\footnote{See also \cite{Alday:2024yyj}.} \cite{Bagchi:2023fbj} at the level of contact correlators, as it was shown that the Carrollian limit of the AdS bulk-to-boundary propagator matches the Carrollian conformal primary wave function which is a modified Mellin transform of a plane wave solution \cite{Banerjee:2018gce}. i.e.,
 \begin{equation}
 	K^{\mp}_{\Delta}(x,u,\theta_p,\phi_p)=\mathcal{N}_\Delta \int_0^\infty d\omega \omega^{\Delta-1}e^{\mp i\omega (u\pm i\epsilon)}e^{i\omega \tilde{q}\cdot x},
 \end{equation}
 where $\mp$ stands for incoming and outgoing respectively. The factor $\mathcal{N}_\Delta$ is a normalisation factor. The flat space limit of AdS bulk-to-bulk propagator was also obtained to be the flat space Green's function, as reviewed in section-\ref{subsec-flat}. To understand how this connects the two approaches, let us consider the $n$-point scalar correlator where, without loss of generality, we take the first two particles as incoming, 
 \begin{align}
 	&C_n(\{p_i,\Delta_i\})\nonumber\\ &= \int d^4 xK^-_{\Delta_1}(x,p_1)K^-_{\Delta_2}(x,p_2) \prod_{i=3}^n K^+_{\Delta_i}(x,p_i)\nonumber\\
 	&=  \int d^4x\prod_i\left( \mathcal{N}_{\Delta_i}\int_0^\infty d\omega_i \omega_i^{\Delta_i-1}\right)e^{-i(\omega_1u_1+\omega_2u_2-\sum_{i=3}^n \omega_iu_i)} e^{-i(\omega_1 \tilde{q}_1+\omega_2 \tilde{q}_2-\sum_{i=3}^n\omega_i \tilde{q}_i)\cdot x}\nonumber\\
 	&=\prod_i\left( \mathcal{N}_{\Delta_i}\int_0^\infty d\omega_i \omega_i^{\Delta_i-1}\right)e^{-i(\omega_1u_1+\omega_2u_2-\sum_{i=3}^n \omega_iu_i)} (2\pi)^4\delta^{(4)}(\omega_1 \tilde{q}_1+\omega_2 \tilde{q}_2-\sum_{i=3}^n\omega_i \tilde{q}_i),
 \end{align}
 where in the last line we have performed the integration over the bulk point to obtain the momentum conserving delta function. We have ignored the $i\epsilon$ prescription as it is not important for the statements here where we are not concerned with the integration contour. Thus, it is clear that the scalar contact Carrollian Witten diagram at $n$ points is the modified Mellin transform of the flat space momentum conserving delta function, which is the corresponding scattering amplitude. It is not hard to generalise this argument to include correlators that contain exchange diagram contributions, by performing Fourier transform of the flat space Green's function, a standard way of going to momentum space in quantum field theory. This involves assuming that the Green's function is translation invariant. As the differential representation is a simple interplay between bulk and boundary translation generators, it also lets us see this easily.
 
 Consider the differential representation obtained in \eqref{flat-diff-rep}. We can see that the differential operators $D_{p_i\mu}$ are diagonalised by the modified Mellin transform kernel since,
\begin{align}
 	D_{p_i\mu} \left(e^{i\omega_i u_i}\right)&=\eta_{\mu\nu}\tilde{q}_i^\nu\frac{\partial}{\partial u_i}\left(e^{i\omega_i u_i}\right),\nonumber\\
 	&=ip_{i\mu}\left(e^{i\omega_i u_i}\right),
 \end{align}
 where $p_{i\mu}=\eta_{\mu\nu}\omega_i\tilde{q}_i^\nu$ is a momentum. Thus we obtain,
\begin{align}
	&\mathcal{A}_n(\{p_i,\Delta_i\})\nonumber\\
	&=\hat{\mathcal{A}}_n(\{D_{p_i\mu}\}) C_n(\{p_i,\Delta_i\}),\nonumber\\
	&=\prod_i\left( \mathcal{N}_{\Delta_i}\int_0^\infty d\omega_i \omega_i^{\Delta_i-1}\right)e^{-i(\omega_1u_1+\omega_2u_2-\sum_{i=3}^n \omega_iu_i)} (2\pi)^4\mathcal{A}_n(\{p_i\})\delta^{(4)}(p_1+p_2-\sum_{i=3}^np_i),
\end{align}
where $\mathcal{A}_n(\{p_i\})$ is the momentum conservation stripped flat space scattering amplitude. Thus we see very transparently that the non-local function of differential operators that resembles the flat space scattering amplitude in the AdS differential representation precisely goes to the flat space scattering amplitude under the modified Mellin kernel\footnote{See related work in the context of dS holography in \cite{Gomez_2021}.}.  While so far in the literature the flat space limit of the differential representation have been discussed in terms of uplifting the flat space momenta to differential operators, and momentum conserving delta functions to contact Witten diagrams at an abstract level, here we see an explicit limit where the differential generators of AdS go to the flat space momentum $4-$vectors under a particular integral transform kernel that diagonalises the differential generators.

 \subsection{Intrinsic conformal Carrollian analysis}
In the previous sections, we have considered the identities satisfied by the AdS bulk-to-boundary propagator and using the flat space limit we obtained the corresponding identities for the Carrollian bulk-to-boundary propagator. Independent of the flat limit of AdS/CFT, we can study the Carrollian holography set up where the conformal Carrollian correlators are defined as/given by the modified Mellin transforms of the flat space scattering amplitudes. The bulk-to-boundary propagator is the Carrollian conformal primary wave function, and the bulk-to-bulk propagator is the flat space Green's function. In this section, we will derive the differential representation using the defining property of the Carrollian conformal primary wave function \cite{Banerjee:2018gce}.

For a conformal primary wavefunction $K_\Delta(x, p)$, corresponding to a bulk Poincar\'e transformation, one can perform a boundary Carrollian conformal transformation such that $K_\Delta (x, p)$ transforms covariantly. For translations, this is given as \cite{Banerjee:2018gce},
\begin{align}
K_\Delta(x+a,u-a\cdot \tilde{q},\theta_{p},\phi_p)=K_\Delta(x,u,\theta_{p},\phi_p),
\end{align}
where $x^\mu$ is the bulk coordinate, $a^\mu$ is the translation parameter, and $\tilde{q}^\mu$ is as given in \eqref{tilde-q}. If we Taylor expand with respect to the translation parameter, we obtain the identities,
\begin{align}
\left(\frac{\partial}{\partial x^\mu}-\eta_{\mu\nu}\tilde{q}^\nu\frac{\partial}{\partial u}\right)K_\Delta(x,p)=0.
\end{align}
 These are clearly the identities that are useful for the differential representation in flat space which we obtained in the previous section by using the flat space limit of AdS/CFT. However, they have a simple origin in the Carrollian framework in terms of the transformation of the Carrollian conformal primary wave function.
 
\paragraph{Celestial case:}
 We discussed earlier that if we take the celestial CFT limit of the AdS differential generators with $\tau=\pm \frac{\pi}{2}$, then we miss out the translation generators. However, in our analysis using Carrollian conformal primary wave function above, what we extracted is the information regarding the translation generators in Carrollian CFT. In the celesetial case, this information is known as well \cite{Stieberger:2018onx,Donnay:2022wvx,Hu:2021lrx} and is given by $i\tilde{q}_\mu e^{\frac{\partial}{\partial \Delta}}$. 
 Therefore, for the celestial conformal primary wave function, we obtain the identity,
 \begin{align}
 	\left(\frac{\partial}{\partial x^\mu}-i\eta_{\mu\nu}\tilde{q}^\nu e^{\frac{\partial}{\partial \Delta}} \right)\Phi_\Delta(x,p)=0.
 \end{align}
 In fact, in the celestial case, the operator $ie^{\frac{\partial}{\partial \Delta_i}}$ acts on the Mellin transform kernel as,
\begin{equation}
ie^{\frac{\partial}{\partial \Delta_i}}(\omega_i^{\Delta_i})=i\omega_i.
\end{equation}
Therefore,
\begin{equation}
i\tilde{q}_{i\mu}e^{\frac{\partial}{\partial \Delta_i}}(\omega_i^{\Delta_i})=ip_{i\mu}(\omega_i^{\Delta_i}).
\end{equation}
Thus even though we do not obtain the translation generators in the celestial case from the flat space limit of the conformal boundary generators in AdS, we can use the prescription \footnote{Such a prescription was also recently advocated in \cite{Ruzziconi:2024zkr} where they construct differential equations for Carrollian MHV gluon and graviton correlators.},
\begin{equation}
D_{p_i\mu}\rightarrow i\tilde{q}_{i\mu}e^{\frac{\partial}{\partial \Delta_i}},
\end{equation}
such that the differential representation carries over to the celestial case.

	\section{Applications}\label{application}
	In this section, we will apply the differential representation for the calculation of Carrollian correlators as well as derive some of their properties. We will illustrate this with the example of 4-point correlators in a scalar $\phi^3$ theory, with the modified Mellin integrals performed. Further, we will show there that the exchange diagams are related to contact diagrams by a differential equation.  In the 4-point example, we will show that this differential equation can be integrated in a simple manner. We will then show another application of the differential representation in showing color-kinematics duality by writing differential Bern-Carrasco-Johannson (BCJ) relations and comment on their relation to the celestial color kinematics duality that exists in the literature \cite{Casali_2021}.

  	\subsection{Calculating exchange Carrollian Witten diagrams from contact diagrams}
	  In the previous section we acted the differential operator under the integral sign which is equivalent to merely saying that in the Fourier space $D_{i\mu}=\omega_i\tilde{q}_{i\mu}$ operators turn out to be momenta. In this section, we consider integrated Carrollian correlators and how their calculation is simplified due to the differential equation provided by the differential representation. Consider the Carrollian differential representation in \eqref{flat-diff-rep}. For simplicity, we shall focus on the example of a four point scalar correlator. For the s-channel differential operator, we have the differential equation,
  	\begin{equation}\label{s-channel-diff-rep}
  		2 {\tilde{q}}_1\cdot{\tilde{q}}_2 \frac{\partial^2}{\partial u_1\partial u_2}\mathcal{A}_4(\{p_i\})=C_4(\{p_i\}).
  	\end{equation}
  	This is a very special partial differential equation. We can solve for the exchange correlator on the left in terms of the contact correlator by integrating the contact correlator first with respect to, say, $u_2$ keeping $u_1$ fixed and then integrate with respect to $u_1$. The integration constants can be fixed by assuming that we are in a generic kinematic region and hence there are no terms which depend only one of the $u_i$'s and not the others. For the case of scalar Carrollian correlators, the contact diagram is a simple rational function in terms of the $u_i$'s and can be easily integrated to obtain the exchange diagram instead of doing the modified Mellin transform all over again to do the calculation. 

    For the higher point correlators, the integration is more complicated since the propagators are generically sums of second order partial derivatives. A generic propagator is of the form,
  	\begin{equation}
  		D_I^2=(D_{p_{i_1}\mu}+D_{p_{i_2}\mu}+\ldots+D_{p_{i_m}\mu})^2,
  	\end{equation}
  	where $I=\{i_1,\ldots,i_m\}$ denote a set of labels. But as we discussed earlier, $D_{p_i,\mu}^2=0$ and hence only cross terms remain. 
  	 \begin{equation}
  		D_I^2=2 \sum_{i_j, i_k \in I} D_{p_{i_j}}\cdot D_{p_{ i_k}}=2\sum_{i_j, i_k \in I}  \omega_{i_j}\cdot \omega_{i_k}\tilde{q}_{i_j}\cdot \tilde{q}_{i_k}\frac{\partial^2}{\partial u_{i_j} 
  			\partial u_{i_k}}.
  	\end{equation}
  	Even though integrating the differential equations with these operators is not as trivial as the four point case, note that all the terms in the sum commute with each other, since $\tilde{q}_{i \mu}$'s don't depend on the $u$-coordinate. Therefore, we can choose to simultaneously diagonalise these differential operators by doing, for instance, a Fourier transform. It will be interesting to see how these differential equations can aid the computation of exchange diagrams in the higher point cases, as well as spinning cases. This we leave for future work. In the next section, we will now get back to the example of 4-point scalar correlators computed in \cite{Bagchi:2023cen} where the exchange and contact correlators were computed separately by performing modified Mellin transform. Here, we will show that the exchange channel for the 4-point correlator can be computed from the contact contribution by integrating \eqref{flat-diff-rep}. 
    
	\subsubsection{Example: 4-point Carrollian correlators in $\phi^3$ theory}\label{after int}
	Consider the scalar contact correlator in $\phi^4$ theory. It was evaluated in \cite{Bagchi:2023cen} to be,
	    \begin{align}\label{4pt-contact}
			C_4(\{p_i,\Delta_i\})
			= & \mathcal{N}_{c} \frac{\delta(|z-\bar{z}|)}{z_{12} \bar{z}_{12} z_{34} \bar{z}_{34}} \prod_{i=1}^4\left((\sigma_i^*)^{\Delta_i-1} \mathbb{1}_{[0,1]}(\sigma_i^*) \right) \nonumber\\ & \times \frac{\Gamma(\sum_{i=1}^4\Delta_i-4)}{ [-i (\sigma_1^* u_1+\sigma_3^* u_3- \sigma_2^* u_2-\sigma_4^* u_4 ) ]^{\Delta_1+\Delta_2+\Delta_3+\Delta_4-4}},
	\end{align}
	where we have stripped away the coupling constant for the interaction and the $\sigma_i^*$'s  are given by \cite{Pasterski:2017ylz},
	\begin{align}
			& \sigma_1^*=-\frac{\epsilon_1 \epsilon_4}{D_4} \frac{z_{24}\bar{z}_{34}}{z_{12}\bar{z}_{13}}, \quad \quad \sigma_2^*=\frac{\epsilon_2 \epsilon_4}{D_4} \frac{z_{34}\bar{z}_{14}}{z_{23}\bar{z}_{12}},\nonumber\\
			& \sigma_3^*=-\frac{\epsilon_3 \epsilon_4}{D_4} \frac{z_{24}\bar{z}_{14}}{z_{23}\bar{z}_{13}}, \quad \quad \sigma_4^*=\frac{1}{D_4}
	\end{align}
	where the denominator $D_4$ is given by,
	\begin{equation}
		D_4=(1-\epsilon_1 \epsilon_4) \frac{z_{24}\bar{z}_{34}}{z_{12}\bar{z}_{13}}+(\epsilon_2 \epsilon_4-1) \frac{z_{34}\bar{z}_{14}}{z_{23}\bar{z}_{12}}+(1-\epsilon_3 \epsilon_4) \frac{z_{24}\bar{z}_{14}}{z_{23}\bar{z}_{13}}.
	\end{equation}
	The	 normalisation factor is given as,
	\begin{equation}
	\mathcal{N}_c=\frac{(-i)^{\Delta_1+\Delta_2}(i)^{\Delta_3+\Delta_4}}{16\pi^6 \prod_{i=1}^4\Gamma(\Delta_i-\frac{1}{2})R^{4-\sum_{i=1}^4\Delta_i}}.
	\end{equation}
	The function $\mathbb{1}_{[0,1]}(\sigma_i^*)$ is $1$ if $\sigma_i^*\in [0,1]$ and vanishes otherwise. Let us try to solve the differential equation \eqref{s-channel-diff-rep} to obtain the exchange Witten diagram using the above. As the contact diagram above is a simple rational function of $u_i$ variables, it is easier to perform the integrate the differential equation. We get,
	\begin{align}
	\mathcal{A}_4(\{p_i\})&=\frac{1}{2\tilde{q}_1\cdot\tilde{q}_2 \partial^2_{u_1u_2}} {C_4(\{p_i\})}\nonumber\\
	&=\mathcal{N}_{c} \frac{1}{2\tilde{q}_1\cdot\tilde{q}_2 z_{12} \bar{z}_{12} z_{34} \bar{z}_{34}}\delta(|z-\bar{z}|) \prod_{i=1}^4\left((\sigma_i^*)^{\Delta_i-1} \mathbb{1}_{[0,1]}(\sigma_i^*) \right) \nonumber\\ & \times \frac{\left(\sigma_1^*\sigma_2^*(\sum_{i=1}^4\Delta_i-4)(\sum_{i=1}^4\Delta_i-5)\right)^{-1}\Gamma(\sum_{i=1}^4\Delta_i-4)}{ [-i (\sigma_1^* u_1+\sigma_3^* u_3- \sigma_2^* u_2-\sigma_4^* u_4 ) ]^{\Delta_1+\Delta_2+\Delta_3+\Delta_4-6}}\nonumber\\
	&=\mathcal{N}_{c} \frac{(1+z_1 \bar{z}_1)(1+z_2 \bar{z}_2)}{4 z_{12}^2 \bar{z}_{12}^2 z_{34} \bar{z}_{34}}\delta(|z-\bar{z}|)(\sigma_1^*)^{\Delta_1-2} (\sigma_2^*)^{\Delta_2-2} (\sigma_3^*)^{\Delta_3-1} (\sigma_4^*)^{\Delta_4-1} \prod_{i=1}^4 \mathbb{1}_{[0,1]}(\sigma_i^*)\nonumber\\ & \times \frac{\Gamma(\sum_{i=1}^4\Delta_i-6)}{ [-i (\sigma_1^* u_1+\sigma_3^* u_3- \sigma_2^* u_2-\sigma_4^* u_4 ) ]^{\Delta_1+\Delta_2+\Delta_3+\Delta_4-6}}.
	\end{align}
	which agrees with the exchange diagram computed by performing modified Mellin integral in \cite{Bagchi:2023cen}. In the last equality above we have used,
	\begin{equation}
		\tilde{q}_i\cdot \tilde{q}_j=\frac{2z_{ij}\bar{z}_{ij}}{(1+z_i\bar{z}_i)(1+z_j\bar{z}_j)}.
	\end{equation} 
	All we had to do is a simple integration of a power law correlator. In addition, we assumed that in a general kinematic region, the correlator would depend on all the $u_i$'s which allowed us to set the integration constants to zero. We can thus see that given the result of the contact Witten diagram one can obtain the exchange diagram by inverting \eqref{s-channel-diff-rep}. This illustrates the use of the differential representation in the computation of Carrollian correaltors.

    \subsection{Differential BCJ relations}
    Color-kinematics duality is an important property of scattering amplitudes in certain theories with color \cite{Bern:2008qj, Bern:2019prr}. One of the applications of differential representation in AdS was to derive color-kinematics duality for AdS correlators \cite{Diwakar:2021juk, Herderschee:2022ntr,Cheung:2022pdk}. The simplest theory that satisfies color-kinematics duality is the bi-adjoint $\phi^3$ theory. In \cite{Diwakar:2021juk} it was shown that the BCJ relations for this theory can be uplifted from flat space to AdS by replacing the Mandelstam variables with differential operators,
    \begin{equation}
    s_{ij}\rightarrow D_i\cdot D_j.
    \end{equation}
    As we have taken the Carrollian limit in this paper, their results carry over to our case, with the replacement,
	\begin{equation}
		s_{ij}\rightarrow 2\tilde{q}_i\cdot\tilde{q}_j \frac{\partial^2}{\partial u_i \partial u_j}.
	\end{equation}
	We expect this to hold in other theories that exhibit color-kinematics duality in flat space as well. In fact, starting with the assumption that boundary Carrollian correlators are modified Mellin transforms of flat space scattering amplitudes, this can be easily argued. For instance, the fundamental BCJ relation for 4-point color-ordered amplitudes in theories with color-kinematic duality reads,
	\begin{equation}
	s_{12}A(1,2,3,4)=s_{13}A(1,3,2,4).
	\end{equation}
	This is nothing but,
	\begin{equation}\label{BCJ-4pt-amp}
		2\tilde{q}_1\cdot\tilde{q}_2 \omega_1\omega_2 A(1,2,3,4)=		2\tilde{q}_1\cdot\tilde{q}_3 \omega_1\omega_3A(1,3,2,4).
	\end{equation}
	Consider the color ordered carrollian correlator that is the modified Mellin transform of the color ordered amplitude,
	\begin{align}
	\mathcal{A}_4(1,\alpha,4)=\prod_{i=1}^4\left(\int d\omega_i \omega_i^{\Delta_i-1}e^{i\omega_i u_i}\right)A(1,\alpha,4).
	\end{align}
	where $\alpha = \{2,3\} \,\text{or}\, \{3,2\}$ denotes the possible orderings and we have ignored the $i\epsilon$ prescription, and considered all particles as outgoing without loss of generality. Clearly,
	\begin{align}
	2\tilde{q}_1\cdot\tilde{q}_2 \frac{\partial^2}{\partial u_1 \partial u_2}\mathcal{A}_4(1,2,3,4)&=
	2\tilde{q}_1\cdot\tilde{q}_2 \frac{\partial^2}{\partial u_1 \partial u_2}\prod_{i=1}^4\left(\int d\omega_i \omega_i^{\Delta_i-1}e^{i\omega_i u_i}\right)A(1,2,3,4)\nonumber\\
	&=\prod_{i=1}^4\left(\int d\omega_i \omega_i^{\Delta_i-1}e^{i\omega_i u_i}\right)(-2	\tilde{q}_1\cdot\tilde{q}_2 \omega_1\omega_2 )A(1,2,3,4)\nonumber\\
	&=\prod_{i=1}^4\left(\int d\omega_i \omega_i^{\Delta_i-1}e^{i\omega_i u_i}\right)(-2	\tilde{q}_1\cdot\tilde{q}_3 \omega_1\omega_3 )A(1,3,2,4)\nonumber\\
	&=2\tilde{q}_1\cdot\tilde{q}_3 \frac{\partial^2}{\partial u_1 \partial u_3}\mathcal{A}_4(1,3,2,4),
	\end{align}
	where the derivatives with respect to $u_i$ on the modified Mellin transform kernel brought down $i\omega_i$ factors, and in the penultimate equality we used the color kinematics duality of the flat space amplitude in \eqref{BCJ-4pt-amp}. Note that this would hold for any theory with color-kinematics duality for its scattering amplitude. This is akin to the celestial case in \cite{Casali_2021} where one uses $ie^{\frac{\partial}{\partial \Delta_i}}$ instead of $\frac{\partial}{\partial u_i}$.

	\section{Discussion}\label{Discussion}
 	In this manuscript, we have considered the flat space limit of the differential representation of AdS correlators. In particular, we performed an In\"on\"u-Wigner contraction of the bulk isometry and boundary conformal differential generators in AdS/CFT to their flat space counter parts. We found that translation generators on the boundary were naturally recovered in the Carrollian case. As the translation generators are of the leading order in the large AdS radius limit, the differential representation for the AdS Laplacian gave rise to the corresponding representation for the flat space Laplacian. Even though such an In\"on\"u-Wigner contraction was suggested in \cite{Li:2023azu} for the bulk isometry generators, it was not extended to boundary conformal generators in a precise flat space holography prescription. 
 	
 	While the resemblance of the differential representation of AdS correlators to flat space scattering amplitudes was appreicated and inspired its applications, a precise limit to boundary correlators in flat space remained elusive. We have filled this gap by showing that the non-local function of differential operators that acts on the contact Witten diagram in the differential representation goes to the flat space amplitude in the large AdS radius limit under the modified Mellin transform kernel. This adds to the harmony between the two prescriptions for Carrollian holography in \cite{Bagchi:2022emh,Bagchi:2023cen}. Further, we have also derived the differential representation intrinsically in Carrollian holography by using the transformation of the Carrollian conformal primary wave function under translations. This elucidates that we are merely using symmetry properties of the Carrollian conformal primary wave functions. Therefore the Carrollian differential representation carries over to other theories beyond the set of theories considered in the AdS differential representation.
 	
 	In our study, we see that the differential equations provided by the differential representation are especially simple in flat space. We used this to show how an exchange channel 4-point scalar Carrollian correlator can be calculated from a contact one by integrating these differential equations.  It will be interesting to see how such calculations can be done for the higher point and spinning correlators. In theories with color, we discussed how the flat space limit of BCJ relations for AdS correlators gives rise to differential BCJ relations for the Carrollian correlator. We also made connection with the celestial case by providing a prescription for the celestial differential representation, which also makes contact with earlier work on celestial BCJ relations \cite{Casali_2021}.

    While computing AdS correlators, the differential representation was used often in conjunction with the split representation of AdS correlators \cite{Costa:2014kfa,Bekaert:2014cea,Eberhardt:2020ewh,Herderschee:2021jbi}. In celestial holography, the split representation of celestial correlators was recently developed in \cite{Chang:2023ttm}. In their work, they foliated the Minkowski spacetime with $EAdS_3$ and $dS_3$ slices. They performed harmonic analysis on these homogeneous spaces to obtain the split representation, which was then uplifted to the celestial case. It will be interesting to see if such an uplift is possible in the context of Carrollian holography. We leave this for future work.
    
 \section*{Acknowledgements}
  We thank Prateksh Dhivakar, Alok Laddha and Prashanth Raman for insightful discussions. SH \& AM thank the organisers of Student Talks on Trending Topics in Theory (ST4) 2024 at IIT Mandi for facilitating discussions. SH thanks IIT Ropar, IIT Bombay, Indian Institute of Science and Harish-Chandra Research Institute for hospitality. SH is funded by the European Union (ERC, UNIVERSE PLUS, 101118787). Views and opinions expressed are however those of the authors only and do not necessarily reflect those of the European Union or the European Research Council Executive Agency. Neither the European Union nor the granting authority can be held responsible for them.
  SC acknowledges ANRF
grant CRG/2022/00616.
  AM would like to thank the Council of Scientific and Industrial Research (CSIR), Government of India, for the financial support through a research fellowship (File No.: 09/1005(0034)/2020-EMR-I). 

\bibliography{main}
\bibliographystyle{JHEP}

\end{document}